\documentclass[twocolumn,showpacs,amsmath,floatfix,prl,aps]{revtex4}
\usepackage{graphicx}

\begin{document}
\title{ Why magnetism in CeO$_{1-x}$F$_x$FeAs and LaO$_{1-x}$F$_x$FeAs is
different }
\author{S. Sharma$^{1,2,3}$}
\email{sangeeta.sharma@physik.fu-berlin.de}
\author{S. Shallcross$^4$}
\author{J. K. Dewhurst$^{1,2,3}$}
\author{A. Sanna$^{2,6}$} 
\author{C. Bersier$^{1,2,3}$}
\author{H. Glawe$^{2,3}$}
\author{E. K. U. Gross$^{2,3}$}
\affiliation{1  Fritz Haber Institute of the Max Planck Society, Faradayweg 4-6, 
D-14195 Berlin, Germany.}
\affiliation{2 Institut f\"{u}r Theoretische Physik, Freie Universit\"at Berlin,
Arnimallee 14, D-14195 Berlin, Germany}
\affiliation{3 European Theoretical Spectroscopy Facility (ETSF)} 
\affiliation{4 Lehrstuhl f\"ur Theoretische Festk\"orperphysik,
Staudstr. 7-B2, 91058 Erlangen, Germany.}

\begin{abstract}
Using state-of-the-art first-principles calculations we study the magnetic
behaviour of  CeOFeAs. We find the 
Ce layer moments oriented perpendicular to those of the Fe layers. An analysis
of incommensurate magnetic structures reveals that the Ce-Ce magnetic coupling 
is rather weak with, however, a strong Fe-Ce coupling. Comparison of 
the origin of the tetragonal to orthorhombic structural distortion in CeOFeAs 
and LaOFeAs show marked differences; in CeOFeAs the distortion
is stabilized by a lowering of spectral weight at the
Fermi level, while in LaOFeAs by a reduction in magnetic
frustration. Finally, we investigate the impact of electron
doping upon CeOFeAs and show that while the ground state Fe moment
remains largely unchanged by doping, the stability of magnetic
order goes to zero at a doping that corresponds well to
the vanishing of the N\'eel temperature.

\end{abstract}

\maketitle

%%%%%%%%%%%%%%%
% Introduction
%%%%%%%%%%%%%%%
The recently discovered \cite{kamihara,ren} family of FeAs based compounds that, 
upon electron
doping, become superconducting with transition temperatures up to 55K, 
are attracting a lot of interest. Structurally
these materials ROFeAs (R=La, Ce, Pr, Nd, Sm) are very similar in that they are 
formed from FeAs layers separated by rare earth or lanthanide oxide layers. 
In striking contrast 
to the well known cuprates, these materials are weakly magnetic metals,
leading to the possibility of large spin fluctuation effects \cite{mazin2,mazin3}.
In particular, at the onset of superconductivity the moment on the
Fe atoms vanishes, and a key question concerns the possible role
of such spin fluctuations in the superconducting transition\cite{mazin2,mazin3}.

Despite the diverse set of rare earth and lanthanide atoms involved
in these materials, physically they share many similarities. 
In particular, (i) at temperatures around $\sim$150K a structural
phase transition from tetragonal to orthorhombic crystal symmetry
occurs and, (ii) this is then closely followed by a
magnetic phase transition to a spin order anti-ferromagnetic (AFM) in nature,
(iii) upon doping with Florine the AFM order is suppressed and 
superconductivity appears.
Furthermore, calculations reveal that (iv) the non-magnetic Fermi 
surfaces are all strongly nested\cite{yin,mazin2,pourovskii}, and 
(v) that the moment of
the Fe atoms depends critically upon the separation of the Fe layer
from the adjacent As layer\cite{yin,mazin1,sharma}.

Given the diversity of constituents involved in this class of materials,
such uniformity of behaviour appears, at first sight, somewhat
surprising. The question then arises if the underlying physical
mechanisms behind such phenomenon as the structural distortion 
or doping behaviour are also the same. In the present work we
investigate this by analyzing the properties of CeOFeAs as
compared to the well studied LaOFeAs. The choice of these two
materials is motivated by the fact that amongst the rare earth and
lanthanide oxypnictides they show relatively large differences
for two key physical properties; (i) the low temperature                 
moment of the Fe atom, 0.35$\mu_B$ in LaOFeAs is the lowest 
while in CeOFeAs (0.94$\mu_B$) is the highest amongst all
oxypnictides\cite{nomura,klauss,cruz,luetkens} , and 
(ii) the superconducting transition temperature also differs strongly; 
26 K in LaOFeAs\cite{kamihara,wen} and 45 K in CeOFeAs\cite{zhao,chen,zocco}.

Remarkably, we find that the mechanism behind the structural
phase transition in CeOFeAs and LaOFeAs is quite different; 
the former case being driven, essentially, by the one electron kinetic
energy i.e. Fermi surface effects, while the latter is driven 
by magnetic frustration\cite{sharma,yildirim}. In addition, the behaviour upon
electron doping is substantially different. We find that
for CeO$_{(1-x)}$F$_x$FeAs the moment is \emph{almost unchanged}
upon electron doping (a reduction of 3\% for $x=0.10$), 
whereas calculations of LaO$_{(1-x)}$F$_x$FeAs
have revealed a strong suppression of the moment (90\% at $x=0.10$).
This may be reconciled with the vanishing of the
N\'eel temperature at $x=0.06$ by the fact that the magnetic order
becomes \emph{meta-stable} near this doping.

%%%%%%%%%%%%%%%%%%%%%%%%%%%%%%
%% Details of calculations  %%
%%%%%%%%%%%%%%%%%%%%%%%%%%%%%%

In the present
work all calculations are performed using the state-of-the-art full-potential
linearized augmented plane wave (FPLAPW) method \cite{Singh}, implemented 
within the Elk code \cite{exciting}.
To obtain the Pauli spinor states, the Hamiltonian containing
only the scalar potential is diagonalized in the LAPW basis: this is the
first-variational step. The scalar states thus obtained are then used as
a basis to set up a second-variational Hamiltonian with spinor degrees
of freedom\cite{Singh}. This is more efficient than simply
using spinor LAPW functions, but care must be taken to ensure that there is
a sufficient number of first-variational eigenstates for convergence of
the second-variational problem. We use a shifted ${\bf k}$ mesh of
$10\times10\times6$, and 260 states per ${\bf k}$-point  
which ensures convergence of the second variational step as well
as the convergence with respect to the ${\bf k}$-points. All the experimental
lattice parameters from Ref. \onlinecite{zhao}  have been used.
                                                       
%%%%%%%%%%%%%%%%%%%%%%%%%%%%%%%%%%%%%%%%%%%%%%%%%%                         
%% Ground state crystal and magnetic structure  %%
%%%%%%%%%%%%%%%%%%%%%%%%%%%%%%%%%%%%%%%%%%%%%%%%%%
\begin{figure}[ht]
\vspace{2cm}
\centerline{\includegraphics[width=0.8\columnwidth,angle=0]{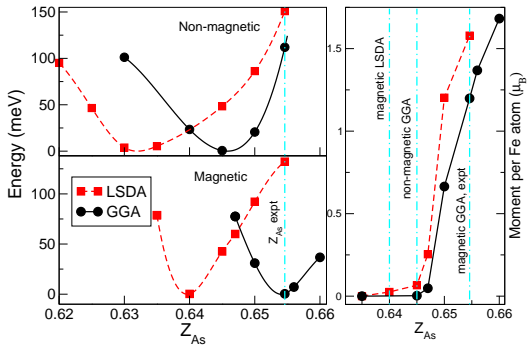}}
\caption{(color online) Top-left panel: energy (in meV) per formula 
unit as a function of position of the As atom calculated using LDA and GGA. Bottom-
left panel: same as upper panel but the calculation is spin-polarized. Right-hand
panel shows the moment per Fe atom (in $\mu_B$) calculated using LSDA and GGA. }
\label{zasvem}
\end{figure}
We first consider the undistorted undoped ground state of CeOFeAs.
One of the most striking aspects of the theoretically intensively studied 
compound LaOFeAs has been the spread of results for magnetic properties,
attributed to both an unusual sensitivity to the approximation to 
exchange-correlation, and a sensitive dependence upon the separation 
of the Fe and As layers, $z_{As}$\cite{yin,mazin1,sharma}. This latter 
behaviour is also found
in the case of CeOFeAs\cite{jishi}, however in contrast to LaOFeAs a spin
polarized generalized gradient approximation (GGA)\cite{pbe} calculation 
yields an optimized $z_{As}$ in near perfect agreement with experiment,
see Fig.~\ref{zasvem}.

Given this choice of $z_{As}$ we now determine the ground state               
magnetic structure using, in addition to the GGA functional,
the local spin density approximation (LSDA)\cite{pw} and LSDA+U\cite{ldapu} 
functionals (for the latter we use $U=6$~eV and $J=1$~eV for the Ce atom).
Considering first only collinear structures, we find that the Fe
layer adopts the stripe AFM structure found
in all the oxypnictides, with the magnetic moments of the Ce layer 
perpendicular to those of the Fe layer. The Fe (Ce) moments are found
to be 1.58$\mu_B$ (0.56$\mu_B$), 1.52$\mu_B$ (0.92$\mu_B$), and 1.30$\mu_B$
(0.54$\mu_B$) for the LSDA, LSDA+U, and GGA functionals respectively,
with the corresponding experimental values 0.94$\mu_B$ for Fe and
0.83$\mu_B$ for Ce. The agreement with experiment for the Fe moment is
thus rather poor for all functionals considered. This may be
attributed to the exceptionally sensitive dependence of the Fe moment on
$z_{As}$, see Fig.~\ref{zasvem}, which for CeOFeAs is even 
more pronounced than in LaOFeAs. Interestingly, as indicated in
experiments\cite{zhao} we find that the Ce and Fe moments
are strongly coupled; a small change (1\%) of the Ce moment (performed
with a fixed spin calculation), leads to a large change (3\%) in the
Fe moment.         
                
%%%%%%%%%%%%%%%%%%%%%%%%%%%
%%    Non-collinear      %%
%%%%%%%%%%%%%%%%%%%%%%%%%%%                                                                            
\begin{figure}[ht]
\vspace{1cm}
\centerline{\includegraphics[width=0.8\columnwidth,angle=0]{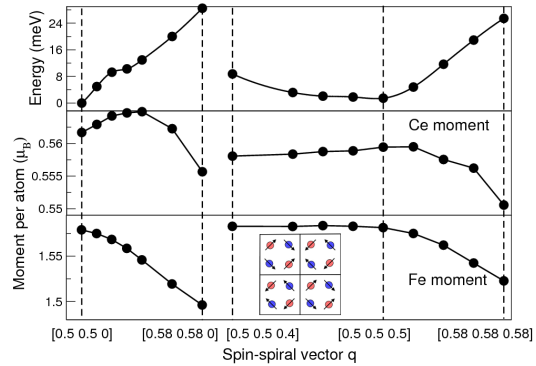}}
\caption{(color online) Top panel shows the energy (in meV per formula unit)
and the middle and bottom panels show the magnetic moment (in $\mu_B$)
per Ce and Fe atom respectively. All quantities are plotted as a function of the
spin-spiral $q$-vector. The inset is a schematic of the spin 
configuration of the Fe and Ce atoms.}
\label{qve}
\end{figure}
In order to study the possible incommensurate spin structure we have
calculated the total energy as a function of the spin-spiral vector {\bf q},
for various directions in the Brillouin zone (BZ).                        
For the in plane spin spirals (going in the direction [0.5, 0.5, 0] to [1, 1, 0])
we find a clear sharp minimum at the commensurate {\bf q} vector of [0.5, 0.5, 0],
equivalent to stripe AFM spin configuration (top panel Fig.~\ref{qve}). 
Interestingly, however, we 
find that the various spin configurations are almost degenerate in the 
direction [0.5, 0.5, 0.4] to [0.5, 0.5, 0.5]. This is indicative of weak
inter-plane Ce-Ce and Fe-Fe coupling; this finding is concomitant with 
experimental results of Zhao \emph{et al.}\cite{zhao}

Turning now to the spin
spiral moments (lower two panels Fig.~\ref{qve}), we find that the
moment remains almost unchanged upon changing the magnetic structure away
from the stripe AFM spin configuration. This is in striking contrast to
LaOFeAs, which remains spin polarized \cite{sharma,yaresko,lorenzana} only 
in a small region about the $M$ point ([0.5, 0.5, 0]), with the moment 
vanishing elsewhere in the BZ.
Since for CeOFeAs the moment on the Fe
atoms changes only slightly - 5.2\% - upon moving across the BZ hence 
the role of spin fluctuations could be rather different in these
two materials.

This difference may be understood as a consequence of the nature
of the various \emph{inter}-plane and \emph{intra}-plane magnetic couplings 
in this material. In particular, the intra-plane coupling of the
Ce atoms is very weak - we find the energy difference between 
FM and AFM ordered in-plane Ce moments to be almost degenerate with,
additionally, the magnitude of the Ce moments unchanged by this choice of FM 
or AFM order. Thus the nature
of the Ce-Ce interaction results in the Ce moment remaining unchanged
by a spin wave configuration; this in turn acts to preserve the Fe moment
due to the strong inter-plane Fe-Ce coupling.
       
%%%%%%%%%%%%%%%%%%%%%%%%%%%
%%      Distortion       %%
%%%%%%%%%%%%%%%%%%%%%%%%%%%
\begin{figure}[ht]
\vspace{1cm}
\centerline{\includegraphics[width=0.8\columnwidth,angle=0]{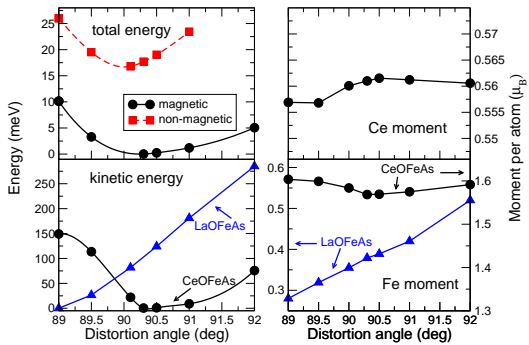}}
\caption{(color online) Top-left panel shows the total energy and bottom-left 
panel the one-electron kinetic energy (in meV) per formula unit. The right panel 
shows the moment per Ce (top) and Fe (bottom) atom (in $\mu_B$) as a function 
of distortion angle (in degrees).}
\label{dvem}
\end{figure}
As with all the oxypnictides, CeOFeAs undergoes a structural phase
transition from tetragonal to orthorhombic crystal 
symmetry\cite{zhao}, which
in this compound occurs at 160K. In order to determine the physical
reason behind this transition we have performed \emph{ab-intio} LSDA calculations.
As may be seen in Fig.~\ref{dvem} the non-magnetic compound does not
show any crystal distortion, but upon performing spin polarized
calculations one finds a minimum at a distortion angle of $90.30^\circ$.
The difference in energy between the non-magnetic undistorted
and magnetic distorted materials is 17meV (197 K), which is 
comparable to the experimental critical temperature of 160K.
Thus both the distortion and critical temperatures are very similar
to those found in LaOFeAs however, as we will now show, the underlying
mechanism of the transition is entirely different.

\begin{figure}[ht]
\vspace{2cm}
\centerline{\includegraphics[width=0.8\columnwidth,angle=0]{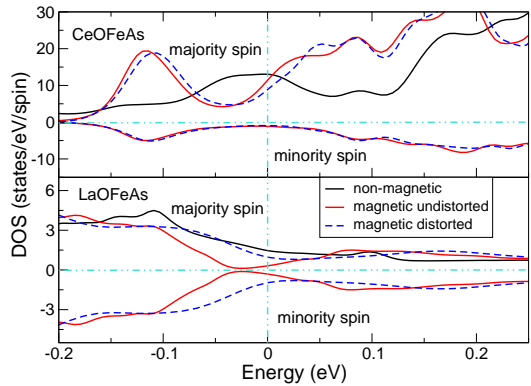}}
\caption{(color online) Density of states calculated using the LSDA for 
undistorted non-magnetic,
undistorted magnetic and distorted magnetic structures for CeOFeAs and LaOFeAs.
Stripe anti-ferromagnetic order was used for the magnetic calculations.}
\label{dos}
\end{figure}

In the case of LaOFeAs the crystal distortion could be rationalized
as being driven by an increase in moment upon distortion\cite{sharma}, due to a
lowering of magnetic frustration in the distorted lattice\cite{sharma,yildirim}. 
On the
other hand, the moment of both Fe and Ce atoms remains essentially
unchanged in CeOFeAs upon distortion, see right hand panels 
of Fig.~\ref{dvem}, thus ruling out such a mechanism for this compound. 
However, the one-electron kinetic energy of CeOFeAs shows
a clear minimum at $90.30^\circ$ while, in dramatic contrast,
the LaOFeAs one electron kinetic energy monotonically increases as a function
of distortion angle. Thus the structural distortion in CeOFeAs is driven
by an electronic structure mechanism, in contrast to the frustration
driven mechanism of LaOFeAs. 
In this regard in Fig. \ref{dos} are shown the non-magnetic as well as magnetic 
(AFM stripe phase) density of states (DOS) for the undistorted and distorted 
lattices.
%In order to validate this hypothesis we 
%show the density of states (DOS) for CeOFeAs and compare it with that 
%of LaOFeAs.
%
One feature common to both compounds
is that, as one would expect, spin polarization results in a large
shift of spectral weight away from the Fermi level. 
Turning to the impact of distortion one observes a striking
difference: in LaOFeAs there is a substantial \emph{increase}
in spectral weight at the Fermi level\cite{yildirim,dong} while, 
in contrast, in CeOFeAs spectral 
weight is moved away from the Fermi level. Thus in this latter case
the redistribution of spectral weight drives the transition, while for 
LaOFeAs the opposite is the case. Clearly, therefore,
although both the distortion angle and transition temperatures
are very similar in these two oxypnictides, the origin of the 
structural phase transitions is very different. At this point it is also worth
mentioning that heat capacity measurements of McGuire \emph{et al.} showed 
that this decrease of the DOS at the  Fermi level in CeOFeAs on distortion 
is consistent with the Seebeck coefficient measurements\cite{mcguire}.

%%%%%%%%%%%%%%%%%%%%%%
%%      Doping      %%
%%%%%%%%%%%%%%%%%%%%%%
\begin{figure}[ht]
\vspace{1cm}
\centerline{\includegraphics[width=0.8\columnwidth,angle=0]{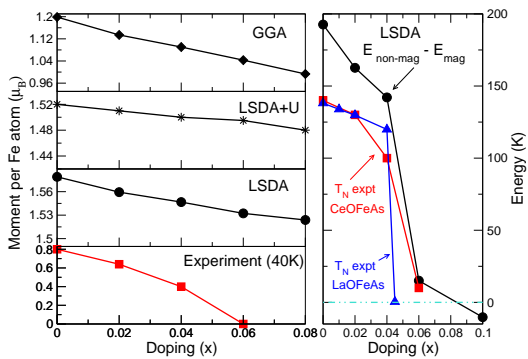}}
\caption{(color online) Left panel: moment per Fe atom (in $\mu_B$) as 
a function of doping calculated using the LSDA, LSDA+U and GGA. The right panel:  
shows the difference between the magnetic and the non-magnetic total energy 
(in K) calculated using the LSDA. The experimental data for CeO$_{1-x}$F$_x$FeAs
are taken from Ref. \onlinecite{zhao} and for LaO$_{1-x}$F$_x$FeAs from 
Ref. \onlinecite{luetkens}}
\label{mvdop}
\end{figure}
The most important property of the iron oxypnictides is the
occurrence of a superconducting phase transition at a critical electron 
doping\cite{zhao,bondino,chi,chen}. An interesting difference
between the temperature-doping phase diagrams of 
CeO$_{1-x}$F$_x$FeAs and LaO$_{1-x}$F$_x$FeAs
is that, for the former case, the N\'eel temperature of the magnetic
phase goes continuously to zero\cite{zhao} as critical doping is
approached ($x_c = 0.06$), while in
LaO$_{1-x}$F$_x$FeAs one instead finds a sharp drop\cite{luetkens,huang,garcia}
in the N\'eel temperature at a critical doping of $x_c = 0.045$ 
(see right hand panel Fig.~\ref{mvdop}). Concomitantly, low temperature
measurements of the magnetic order of the Fe layer in CeO$_{1-x}$F$_x$FeAs show
that it is entirely lost before $x_c$ (left panel Fig.~\ref{mvdop}).
An important question for understanding the role of magnetism
in the superconducting transition is then whether it is the
moment of the Fe atoms that vanishes, or whether it is simply
the stripe AFM order that vanishes.

To investigate this issue we calculate CeO$_{1-x}$F$_x$FeAs
by deploying the virtual crystal approximation (VCA). 
Although the VCA neglects much of the physics
of disorder, it has been shown recently\cite{larson} to provide
a surprisingly accurate account of magnetism and Fermiology in
LaO$_{1-x}$F$_x$FeAs. In order to determine the impact of
the approximation to the xc functional, calculations were
performed using the LSDA, LSDA+U, and
GGA functionals.

Amongst the undoped Fe oxypnictides the Fe moment
is largest in CeOFeAs and smallest in LaOFeAs, and it is therefore
surprising that in the latter compound the magnetic phase
appears more stable for $x<x_c$.
Indeed, calculations of the Fe moment in
CeO$_{1-x}$F$_x$FeAs (left panel Fig.~\ref{mvdop}) show that,
in apparent contradiction to the experimental data, 
there is \emph{almost no change} upon doping. However,
if we consider the magnetization energy, i.e. the quantity
$E_{\text{non-mag}} - E_{\text{mag}}$ (right panel Fig.~\ref{mvdop}), 
this falls sharply at exactly the critical doping $x_c = 0.06$.
Thus it is the \emph{stability} of the magnetic order which
is reduced upon doping, not the actual Fe moment. This sudden
reduction of stability, which is not found in 
LaO$_{1-x}$F$_x$FeAs, is behind the differing forms of the
N\'eel temperature phase boundary.

In the case of LaOFeAs calculations have shown\cite{sharma} that, upon doping,
the distortion angle changes from an undoped value of 
$\theta=90.27^\circ$ to $\theta=89.85^\circ$; additionally, there
is the appearance of incommensurate magnetic phases. We have
investigated if such phases appear in the doping phase diagram 
of CeO$_{1-x}$F$_x$FeAs and find that (i) the distortion angle is unchanged 
upon doping ($0 \le x \le 0.06$) and (ii) the ground state magnetic structure 
remains stripe AFM.

%%%%%%%%%%%%%%%
% Conclusions %
%%%%%%%%%%%%%%%
To conclude, we have performed a comparative
study of the two oxypnictides CeO$_{1-x}$F$_x$FeAs
and LaO$_{1-x}$F$_x$FeAs. Although many common features, e.g.
a similar non-magnetic Fermi surface, may be found in
the iron oxypnictides we have shown here that profound
differences also exist. In particular we find that 
the structural distortion in CeOFeAs is driven by one electron kinetic energy
(i.e. electronic structure effect), in contrast to LaOFeAs 
where a lowering of magnetic frustration has been identified
as the mechanism. Furthermore, the behaviour under electron
doping differs markedly between the two materials; in
CeO$_{1-x}$F$_x$FeAs we find that while the ground state
moment is essentially unchanged with doping, the \emph{stability}
of the moment is sharply reduced. Finally, we have -
via calculations of incommensurate spin spiral structures -
carefully investigated the various magnetic couplings 
of the constituents of CeOFeAs; we find that the Ce atoms
are only weakly coupled to each other, while the Fe atoms are much more 
strongly coupled to each other as well as to the Ce.

%\bibliography{ceofeas}
\end{document}